\documentclass[12pt,letterpaper]{article}
\usepackage[a4paper, total={7in, 10in}]{geometry}

\usepackage{graphicx}
\usepackage{authblk}
\usepackage{hyperref}
\usepackage{amsmath} 
\usepackage{amssymb} 
\usepackage[super,comma,sort&compress]  
   {natbib}
\usepackage{float}

\usepackage{booktabs}

\usepackage{helvet}

\makeatletter
\renewcommand{\maketitle}{\bgroup\setlength{\parindent}{0pt}
\begin{flushleft}
  \textbf{\@title}
  
  \@author
\end{flushleft}\egroup}
\makeatother

\title{Estimation of $^{97}$Ru Half-Life Using the Most Frequent Value Method and Bootstrapping Techniques}
\date{\today}

\author[1,*]{Victor V. Golovko}

\affil[1]{Canadian  Nuclear  Laboratories \newline 286 Plant Road, Chalk  River, K0J~1J0, Ontario, Canada}

\affil[*]{Correspondence: victor.golovko@cnl.ca}

\begin{document}

\maketitle

\section*{Abstract}

A new and robust statistics was applied to previous measurements of the $^{97}$Ru half-life. This process incorporates the most frequent value (MFV) technique along with hybrid parametric bootstrap (HPB) method to deliver a more precise estimate of the $^{97}$Ru half-life. The derived value is $T_{1/2\text{,MFV}}(\text{HPB}) = 2.8385^{+0.0022}_{-0.0075}$ days. This estimate corresponds to a 68.27\% confidence interval ranging from 2.8310 to 2.8407 days, and a 95.45\% confidence interval ranging from 2.8036 to 2.8485 days, calculated using the percentile method. This level of uncertainty is significantly lower--over 30 times--than the uncertainty in the previously recognized half-life value found in nuclear data sheets. Employing an alternate approach to minimization could further cut down the statistical uncertainty by 44\% for the $^{97}$Ru half-life. In particular, the HPB method accounts for uncertainties in small datasets when determining the confidence interval. When the HPB method, in combination with the MFV approach, was used to review a four-element dataset of the specific activity of $^{39}$Ar based on underground data, the result was $ SA_{\text{MFV}}(\text{HPB}) = 0.966^{+0.027}_{-0.020}$ Bq/kg$_{\text{atmAr}}$. This value results in a 68.27\% confidence interval of 0.946 to 0.993, along with a 95.45\% confidence interval of 0.921 to 1.029, also determined using the percentile method.

\section*{Keywords}

Most frequent value, Bootstrap analysis,  $^{97}$Ru half-life, $^{39}$Ar specific activity, Hybrid parametric bootstrapping 

\section*{Introduction}
\label{intro}

A common challenge when working with small datasets is how to accurately average data that include outliers. The Particle Data Group (PDG) offers a practical solution for optimal averaging, as detailed in~\cite{10.1093/ptep/ptaa104}. However, some measurements may need to be excluded from the averaging process for various reasons, such as the absence of error values, questionable assumptions, inconsistency with more reliable results, or being derived from preprints or conference reports. Excluding data from small datasets further reduces the sample size, potentially compromising the reliability of the averaging results. Additionally, excluded data might have been obtained via different techniques, each with distinct systematic effects.

In this paper, we propose a robust method that uses the most frequent value (MFV) approach combined with the bootstrapping technique to average small datasets that include outliers. As a case study, we apply these statistical techniques to historical data on the half-life measurements of $^{97}$Ru. This dataset was also chosen to investigate whether the use of different fitting algorithms impacts the fit results and statistical uncertainty.

Furthermore, $^{97}$Ru possesses advantageous physical characteristics, including its half-life, decay mode, and strong $\gamma$-decay emissions, making it suitable for specific types of diagnostic imaging and treatment. $^{97}$Ru is a promising candidate in nuclear medicine~\cite{MaitiLahiri2011_359_364}. Its relatively long half-life allows $^{97}$Ru-labeled compounds to monitor prolonged metabolic processes effectively~\cite{871}. The isotope $^{97}$Ru could  replace $^{99m}$Tc and $^{111}$In in various applications~\cite{Kumar2021}.

In the `\nameref{sec:stat}' section, we provide an overview of the statistical techniques used in this work. The section `\nameref{sec:hisRu97}' reviews the historical data on the half-life of $^{97}$Ru and includes a reanalysis of one decay dataset via a new and updated version of the minimization software. We use this approach to estimate the confidence interval for the $^{97}$Ru half-life. Furthermore, in the subsection `\nameref{subs:boot},' we re-evaluate the small set, which comprises only four elements of the $^{39}$Ar specific activity from underground measurements, via a new technique. 

\section*{Methodology }
\label{sec:stat}

In this section, we briefly discuss the traditional statistical methods used by PDG~\cite{10.1093/ptep/ptaa104} for averaging data and determining its uncertainty. We will outline the basic averaging techniques, specifically focusing on the commonly used weighted least-squares procedure. Additionally, we  provide an overview of the most frequent value method, which is effective for analyzing data that have outliers. Furthermore, we delve into both nonparametric and parametric bootstrap techniques used to estimate the data spread, which in turn help in determining the confidence interval even for  data that are not normally distributed. Finally, we introduce the hybrid parametric bootstrap method suitable for small datasets, as it also considers the uncertainty of each data point.

\subsection*{Weighted average method}
\label{subs:wam}

The PDG offers a detailed method for performing unconstrained fitting or averaging of independent measurements~\cite{10.1093/ptep/ptaa104}. In this context, the uncertainty of each measurement is considered to follow a Gaussian distribution, representing a 68.3\% confidence interval around the central value. Here, we present only the details  related to the historical dataset for the $^{97}$Ru half-life.

Assuming that the measurements of the half-life of $^{97}$Ru are not correlated, the weighted average \( \bar{x} \) for a set of measurements \( x_i \) with their respective uncertainties \( \sigma_i \) across \( N \) experiments is determined by:
\begin{equation}
	\bar{x} = \frac{\sum_{i=1}^{N} \frac{x_i}{\sigma_i^2}}{\sum_{i=1}^{N} \frac{1}{\sigma_i^2}}.
\end{equation}
The uncertainty in the weighted average \( \sigma_{\bar{x}} \) is calculated by:
\begin{equation}
	\sigma_{\bar{x}} = \sqrt{\frac{1}{\sum_{i=1}^{N} \frac{1}{\sigma_i^2}}}.
\end{equation}
It is also recommended to calculate the chi-square \(\chi^2\) value for the series of $^{97}$Ru half-life measurements, which can be expressed as:
\begin{equation}
	\chi^2 = \sum_{i=1}^{N} \left(\frac{x_i - \bar{x}}{\sigma_i}\right)^2
\end{equation}
If $^{97}$Ru half-life measurements follow Gaussian distributions, the expected value of \(\chi^2\) is approximately equal to \(N-1\). The quality of the averaging result is checked via a reduced chi-squared value \(\chi^2/(N-1)\). If \(\chi^2/(N-1)\) is less than or equal to one, then the dataset is considered acceptable, and the weighted average along with its uncertainty is acceptable. If \(\chi^2/(N-1)\) is greater than one, the weighted average is still acceptable, but the weighted uncertainty should be increased by a scaling factor of \( \sqrt{\chi^2/(N-1)} \). Details are explained by Zyla et.al. (2020)~\cite{10.1093/ptep/ptaa104}.

\subsection*{The most frequent value method}

The traditional mode in statistics identifies the most frequent value in a dataset~\cite{illowsky2018introductory}. This is easy when one value is clearly more frequent. However, in datasets where values are spread out and not repeated, the traditional mode may not find a representative central value. The MFV method addresses this by finding a central value even in such cases, minimizing information loss~\cite{zhangMostFrequentValue2017}. Unlike the traditional mode, which relies on frequency counts, the MFV method can identify a meaningful central value in datasets with no predominant value. A practical example using an artificial dataset to show differences in application of the mode and MFV is in~\cite{golovko2023unveiling}.

Steiner introduced a robust statistical model called the MFV, which is outlined in detail in \cite{Steiner1973,Csernyak1973, Ferenczy1988ShortIntroduction, steinerMostFrequentValue1988, steinerIntroductoryInstructionsComputations1991, steinerMostFrequentValue1991, steinerOptimumMethodsStatistics1997, kemp2004steiner}.
To find the most common value in a dataset ($x_1,...,x_i,...,x_N$), we can calculate the MFV ($M$) and the scale parameter (denoted by $\varepsilon$ and called dihesion) iteratively. This task involves calculating a specific mathematical expression multiple times. The expression is as follows:
\begin{equation}
	M_{j+1} = \frac{ \sum_{i=1}^{N} x_i \cdot \frac{\varepsilon^2_j}{\varepsilon^2_j + \left( x_i - M_j \right)^2 } }{ \sum_{i=1}^{N} \frac{\varepsilon^2_j}{\varepsilon^2_j + \left( x_i - M_j \right)^2} }.
	\label{Eq:MFV}
\end{equation}
In this equation, $\varepsilon_j$ is also calculated as a result of an iterative process that is performed a certain number of times. The formula for this calculation is as follows:
\begin{equation}
	\varepsilon^2_{j+1} = 3 \cdot \frac{ \sum_{i=1}^{N} \frac{( x_i - M_j)^2 }{ \left( \varepsilon^2_j + \left( x_i - M_j \right)^2 \right)^2 } }{ \sum_{i=1}^{N} \frac{ 1 }{ \left( \varepsilon^2_j + \left( x_i - M_j \right)^2 \right)^2 } }.
	\label{Eq:dihesion}
\end{equation}

To start the iterations ($j=0$), we first set the initial values. We set $M_{(0)}$ as the mean of the dataset, which is calculated as $\frac{1}{N} \sum_{i=1}^{N} x_i$. 
It is recommended to start with the median $M_{(0)}$ instead of the mean when performing the MFV iterations \cite{HajagosSteiner1992}. This is because the mean can be heavily influenced by outliers or the tails of the distribution, which could lead to many additional iteration steps to reach the desired result. In cases where the initial value of $\varepsilon_j$ is significantly larger than the correct value, the number of iteration steps on the $\varepsilon_j$ branch may need to be increased significantly.

The study by Hajagos \cite{Hajagos1980} discussed a method to find the value of $\varepsilon_{(0)}$ efficiently for rapid calculations of the MFV. In simple terms,  we can set $\varepsilon_{(0)} = \frac{\sqrt{3}}{2} \cdot (x_{\text{max}} - x_{\text{min}})$, where $x_{\text{max}}$ and $x_{\text{min}}$ are the highest and lowest values in the dataset and are used to estimate the MFV. This formula helps in determining $\varepsilon_{(0)}$ accurately for the MFV calculations.
In the process, we need to ensure that both Equations~\ref{Eq:MFV} and \ref{Eq:dihesion}, which are used to compute the values of $M_{j+1}$ and $\varepsilon^2_{j+1}$ from the datasets, are satisfied simultaneously.

The calculation of the empirical value of $M_{j+1}$ can be computationally intensive. Various practical applications of this method are discussed in \cite{FerenczySteiner1988MostFrequentValues,  szucsApplicabilityMostFrequent2006,  zhangMostFrequentValue2017,  szaboMostFrequentValuebased2018, zhang2018most, zhang2022mfv, golovkoApplicationMostFrequent2023, golovko2023unveiling}. 

Csernyak and Steiner \cite{CsernyakSteiner1983} provided a simple formula for calculating the variance $\sigma_{M_j}$ at each iteration step in a symmetrical distribution when $M_{j+1}$ is determined according to Equation~\ref{Eq:MFV}:
\begin{equation}
	\sigma_{M_j} = \frac{\varepsilon_j}{\sqrt{n_{\text{eff}}}}.
	\label{Eq:sig_M}
\end{equation}
In this formula, $\varepsilon$ represents the dihesion (which is the convergence value of the iterations as defined in Equation~\ref{Eq:dihesion}), and $n_{\text{eff}}$ is the effective number of data points. The number of effective points influencing the MFV result~\cite{Csernyak1980, HajagosSteiner1992} is calculated as:
\begin{equation}
	n_{\text{eff}} = \sum_{i=1}^{N} \frac{\varepsilon_j^2}{\varepsilon_j^2 + \left( x_i - M_j \right)^2}.
\end{equation}
This formula offers a way to determine the variance that characterizes the accuracy when  $M$ is estimated for symmetrical distributions. 

\subsection*{Bootstrapping method}
\label{subs:boot}

To understand the uncertainty or variability of a statistical model or specific measure such as the MFV, researchers employ a technique known as bootstrapping, as discussed in Davison's work~\cite{davisonBootstrapMethodsTheir1997}. This method involves repeatedly sampling from the existing dataset by selecting samples with replacement, termed ``bootstrap samples.'' By analyzing these samples, researchers can create a distribution for the statistic of interest, such as the MFV, and therefore, the confidence intervals, which help to predict the range of possible values and thereby make more informed decisions about the data or model in question.

Additionally, by integrating the MFV method with bootstrapping, the precision of field data, including the half-life of $^{97}$Ru, can be improved using existing historical data without resorting to further experiments. Real-world applications have shown that the MFV method can handle outliers effectively~\cite{golovkoApplicationMostFrequent2023,golovko2023unveiling}. Moreover, we will include the historical $^{97}$Ru half-life data noting that, according to the Particle Data Group's recommendations, these data should be excluded since they are inconsistent with other, more reliable results or lack specified errors. By combining the MFV approach, which relies on minimizing information loss, with bootstrapping  techniques, a more effective and unconstrained averaging of the results becomes possible.

Bootstrapping is beneficial whether or not there is a clearly defined probability model for the data~\cite{efron1994introduction,davisonBootstrapMethodsTheir1997}. It allows for the generation of a statistic's distribution, aiding in drawing more accurate conclusions about the studied population or model. There are two forms of the bootstrap method: nonparametric bootstrapping and parametric bootstrapping. Nonparametric bootstrapping involves resampling from the original sample, whereas parametric bootstrapping involves resampling from the model fitted to the original sample~\cite{puth2015variety}. This distinction is important for understanding how the bootstrap method can be applied in different scenarios. 

The bootstrapping process itself is straightforward but requires significant computational power. Importantly, bootstrapping does not compensate for small sample sizes; it does not create new data or fill in missing dataset values. Instead, it provides insights into how additional samples might behave if drawn from a population similar to the original sample~\cite{bruce2020practical}.

However, when the original dataset is small, enhancing confidence interval estimations can be achieved through a simple parametric bootstrap method. This approach assumes that the data follow a specific distribution--often a normal distribution--and does not consider the individual uncertainties in the data measurements. Starting with a hypothesis about the data's distribution, it uses estimated parameters such as the mean and variance from the original data to generate new, synthetic datasets. These datasets are then used to recalculate the desired statistics (for example the mean or confidence interval), and the process is repeated to build a distribution of the estimator.

By reviewing the recalculated statistics' distribution, this method sheds light on the estimator's variance, potential bias, and confidence intervals, offering statistical inference on the basis of the model's assumption. This approach proves invaluable when the data distribution is well understood and can be parametrically defined, facilitating statistical tests and confidence estimations that might otherwise be impractical with direct population sampling.

Here, we  demonstrate how to apply a technique similar to parametric bootstrapping on a tiny artificial dataset containing only four elements  along with their 1$\sigma$ normal uncertainties. This method  helps us better understand the process. Parametric bootstrapping is helpful for estimating confidence intervals, especially with small sample sizes. 
It assumes that we know the probability distribution function of the original dataset, or can deduce its parameters from the data. However, in the method we demonstrate, we do not need to know the exact distribution function of the original dataset. Instead, we have analytical functions that are used to calculate each value in the dataset and its uncertainty. This information comes from the analysis method used to obtain each value in the dataset. This technique, which we call hybrid parametric bootstrapping (HPB), is helpful for handling datasets that have a limited number of data points, each with a known level of uncertainty.

\begin{figure}[t] %
	\centering
	\includegraphics[width=0.9\columnwidth]{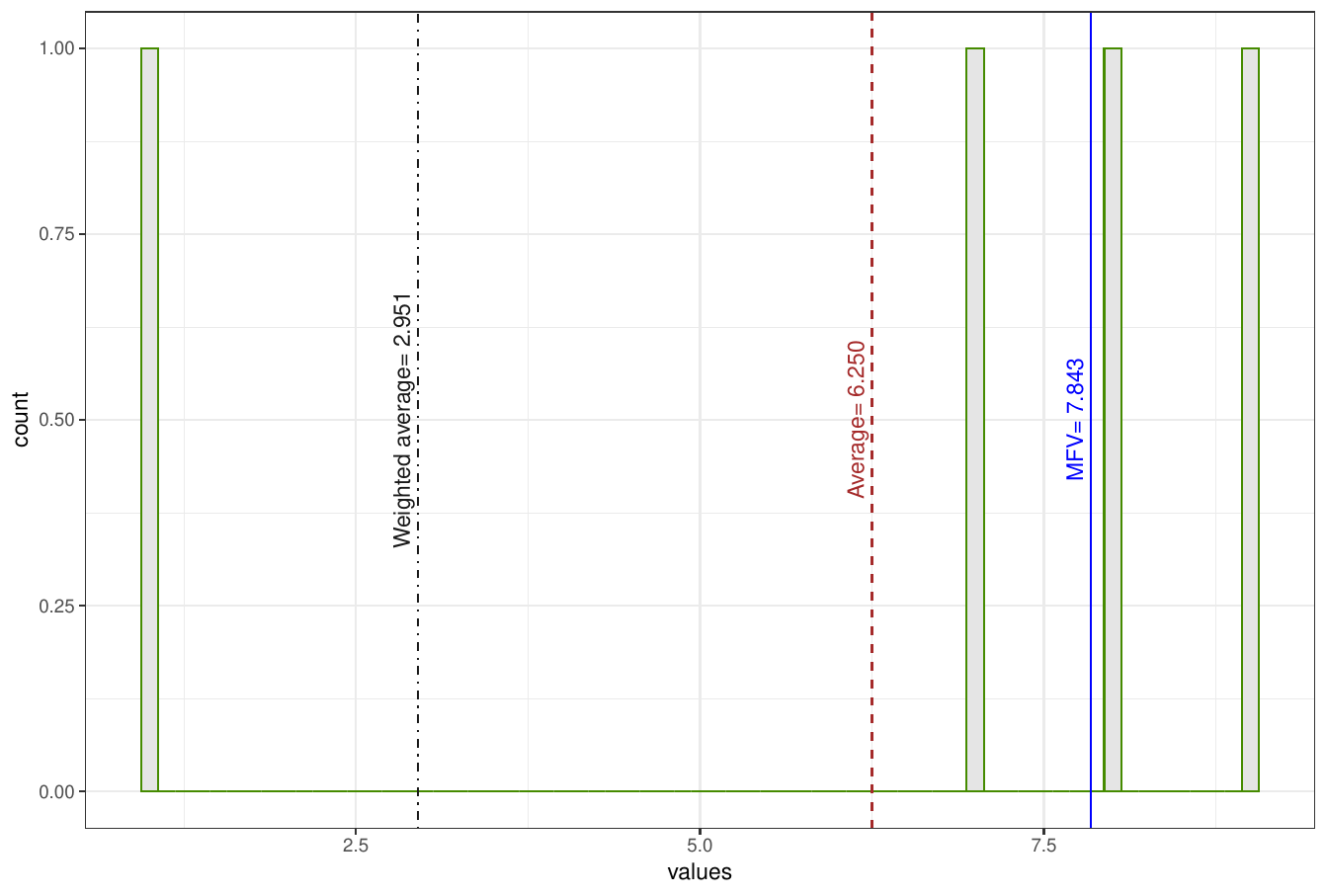}
	\caption{An artificial dataset  composed of four elements. The vertical dashed line shows the mean value, whereas the vertical solid line indicates the MFV. The vertical dotted-dashed line represents the weighted average value.}
	\label{fig:art_dataset}
\end{figure}

Let us consider an artificial dataset consisting of four elements ($x_1, x_2, x_3, x_4$) with theoretical average values of 1, 7, 8, and 9, respectively. Each of these elements is associated with defined uncertainties ($\sigma_1, \sigma_2, \sigma_3$, and $\sigma_4$), represented by hypothetical standard deviations of 0.1, 0.2, 0.3, and 0.4, respectively. 
The histogram in Figure~\ref{fig:art_dataset} displays the distribution of the artificial dataset. This histogram allows us to visually understand the range of values and their frequency of occurrence. The plot also shows the average (6.25), the weighted average (2.95), and the MFV (7.84) for the dataset. Notably, in this artificial dataset, the average value is lower than the MFV, whereas the weighted average is close to the value with the least uncertainty. 

Nonparametric bootstrapping starts by taking an initial sample of data and then repeatedly resampling them from these data. This method is useful because it relies solely on the data in the original dataset and does not assume any specific statistical shape or characteristics of the broader population. However, it does not consider the uncertainty associated with each individual element in the sample. 
Furthermore, it does not need any prior knowledge about the population from which the sample is drawn. Nonparametric bootstrapping is a versatile and powerful method for estimating statistical properties such as point estimates (e.g., mean or MFV) or the distribution's spread (e.g., confidence interval). However, when dealing with a small original dataset, it is advisable to use a parametric bootstrap method. 

\begin{figure}[t] %
	\centering
	\includegraphics[width=0.9\columnwidth]{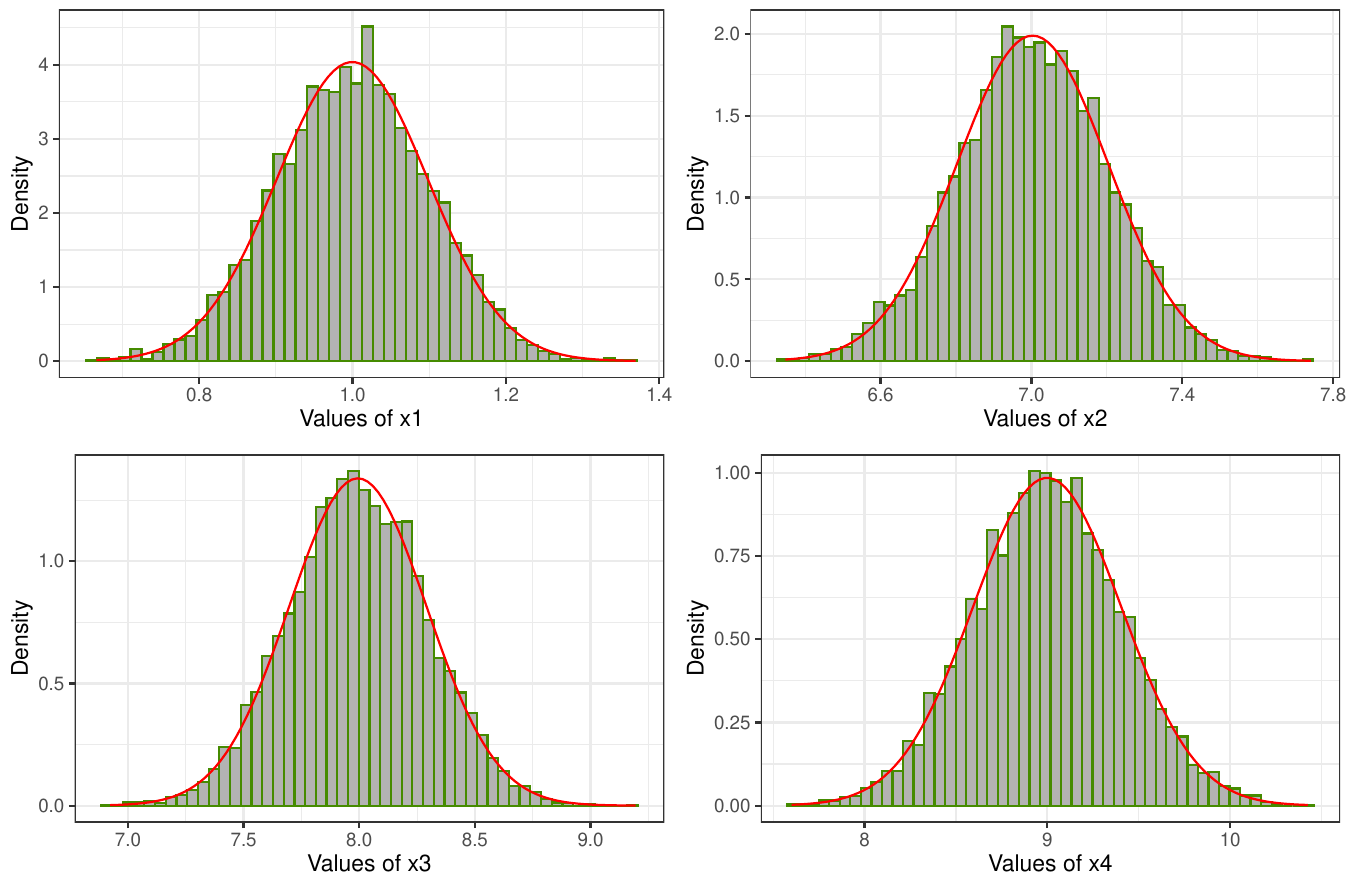}
	\caption{A histogram was created for each element, which was sampled with replacement from the original dataset, which initially had four elements. The histogram also shows the uncertainty associated with each element.}
	\label{fig:bootstrap_ind}
\end{figure}

Both nonparametric and parametric methods do not consider the uncertainty of the elements in the original dataset. Therefore, the following question arises: is it possible to adapt a parametric bootstrap approach, which performs well on small datasets, to incorporate a nonparametric approach of bootstrapping from the original dataset while also considering the uncertainty of each element in the dataset? This modification to parametric bootstrapping  enables us to statistically estimate the variability of the original limited dataset (and therefore to estimate its confidence interval) without prior knowledge of the underlying data population distribution.  On the basis of the measured results,  we already know the statistical model (for example, a normal distribution) that describes how each element's value ($\mu$) in the original dataset and its uncertainty ($\sigma$) were determined.

To predict different results considering the uncertainties of the elements and their values, we apply a bootstrap sampling to each element in the original dataset. This method involves generating a new bootstrap data sample by randomly selecting from normal distributions centered around the mean value of each element in the original dataset and influenced by their uncertainties. By doing this, we can generate artificial data points that represent possible variations that could occur in real-world scenarios on the basis of the provided information. This process allows us to better understand the potential outcomes of the original data and their spread. 

The histograms in Figure~\ref{fig:bootstrap_ind} display the distribution of each element that was generated for bootstrap samples and a normal fit that was performed for each element. Table~\ref{tab:comparison_stats} displays the fitting results obtained from the data generated through bootstrapping for each element in the original dataset. These results were derived via the normal distribution function. Additionally, the table presents the absolute percent difference between the estimated mean and its uncertainty. This information helps in comparing the fit results with the original dataset. 

\begin{table*}[t]
	\centering
	\caption{Comparison of original and estimated means with uncertainties for the artificial dataset.  The absolute percent difference between the estimated mean and its uncertainty is also shown.} 
	\label{tab:comparison_stats}
	\begin{tabular}{ccccccc}
		\toprule
		Element & 
		\begin{tabular}{c}
			Original \\
			$\mu_{\rm{o}}$ \\
		\end{tabular} & 
		\begin{tabular}{c}
			Original \\
			$\sigma_{\rm{o}}$ \\
		\end{tabular} & 
		\begin{tabular}{c}
			Fitted \\
			$\mu_{\rm{f}}$ \\
		\end{tabular} & 
		\begin{tabular}{c}
			Fitted \\
			$\sigma_{\rm{f}}$\\
		\end{tabular} & 
		\begin{tabular}{c}
			$| \frac{\mu_{\rm{f}}\ - \mu_{\rm{o}}}  {\mu_{\rm{o}}} |$,
			\% \\
		\end{tabular} & 
		\begin{tabular}{c}
			$| \frac{\sigma_{\rm{f}}\ - \sigma_{\rm{o}}}  {\sigma_{\rm{o}}} |$, \% \\
		\end{tabular} \\ 
		\midrule
		$x_1$ & 1.0 & 0.1 & 0.9996 & 0.0988 & 0.038 & 1.22 \\ 
		$x_2$ & 7.0 & 0.2 & 7.0030 & 0.2004 & 0.043 & 0.21 \\ 
		$x_3$ & 8.0 & 0.3 & 7.9941 & 0.2980 & 0.074 & 0.67 \\ 
		$x_4$ & 9.0 & 0.4 & 8.9987 & 0.4053 & 0.014 & 1.33 \\ 
		\bottomrule
	\end{tabular}
\end{table*}

\begin{figure}[t] %
	\centering
	\includegraphics[width=0.9\columnwidth]{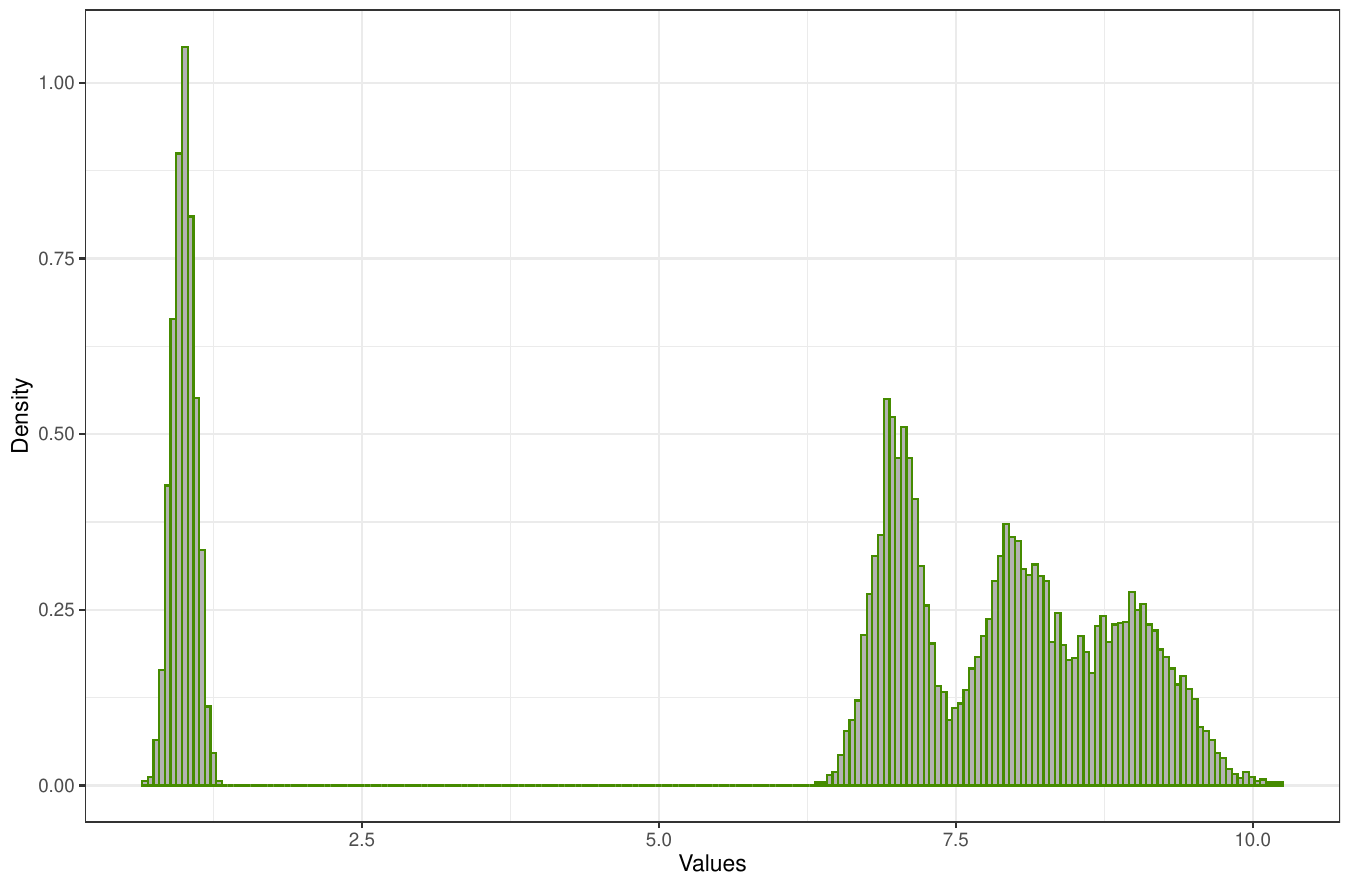}
	\caption{A histogram of all the bootstrap data from artificial datasets that contain four elements.}
	\label{fig:bootstrap_all}
\end{figure}

Figure~\ref{fig:bootstrap_all} shows all bootstrap datasets that were obtained via the HPB approach, as histogram. This hybrid method enhances the ability to estimate variability and confidence intervals in small datasets, which is especially useful in fields such as environmental remediation, where data points may be limited and precise modeling of uncertainty is crucial. For the histogram a total of 10,000 bootstrap datasets were generated, each having four elements. For each bootstrap dataset the MFVs were calculated via Equation~\ref{Eq:MFV}.

A key aim of this study is to create the most cautious approach for applying statistical techniques to analyze data from small historical datasets, such as half-life measurements. We specifically aim to estimate the confidence interval for a small dataset, considering both the data values and their associated uncertainties.

We have introduced the HPB method, which is designed specifically for small datasets that may not follow traditional distribution assumptions. Rather than assuming an unknown distribution, the HPB method uses analytical functions that were initially employed to compute each data point along with its uncertainty. This method proves to be highly beneficial when dealing with small datasets, as demonstrated by its application to an artificial dataset comprising only four elements. 

We can evaluate the HPB method via a small dataset that includes four specific activity measurements conducted in different underground laboratories through various experiments (refer to Table~\ref{tab:SA_Ar39}). This dataset was analyzed via nonparametric bootstrapping with the MFV approach~\cite{golovkoApplicationMostFrequent2023}. It would be interesting to compare the confidence interval obtained by nonparametric bootstrapping to that obtained via the HPB method.

\begin{table}[t]
	\centering
	\caption{Summary of specific activity ($SA$) measurements of $^{39}$Ar in the atmosphere from different underground experiments.}
		\begin{tabular}{lccc}
			\toprule
			Experiment &  \begin{tabular}{c}
				$SA$,\\
				Bq/kg$_{\text{atmAr}}$\\
			\end{tabular} & \begin{tabular}{c}
				Uncertainty,\\
				Bq/kg$_{\text{atmAr}}$\\
			\end{tabular} & Ref.\\
			\midrule
			DEAP-3600 & 0.964 & $ \pm $0.024 & \cite{adhikariPrecisionMeasurementSpecific2023a} \\
			DEAP-3600 & 0.97 & $ \pm $0.03 & \cite{adhikariPrecisionMeasurementSpecific2023a} \\
			WARP & 1.01 & $ \pm $0.08 & \cite{benetti2007measurement} \\
			ArDM  & 0.95  & $ \pm $0.05 & \cite{calvo2018backgrounds} \\
			\bottomrule
		\end{tabular}%
	\label{tab:SA_Ar39}%
\end{table}%

\begin{figure}[t] %
	\centering
	\includegraphics[width=0.9\columnwidth]{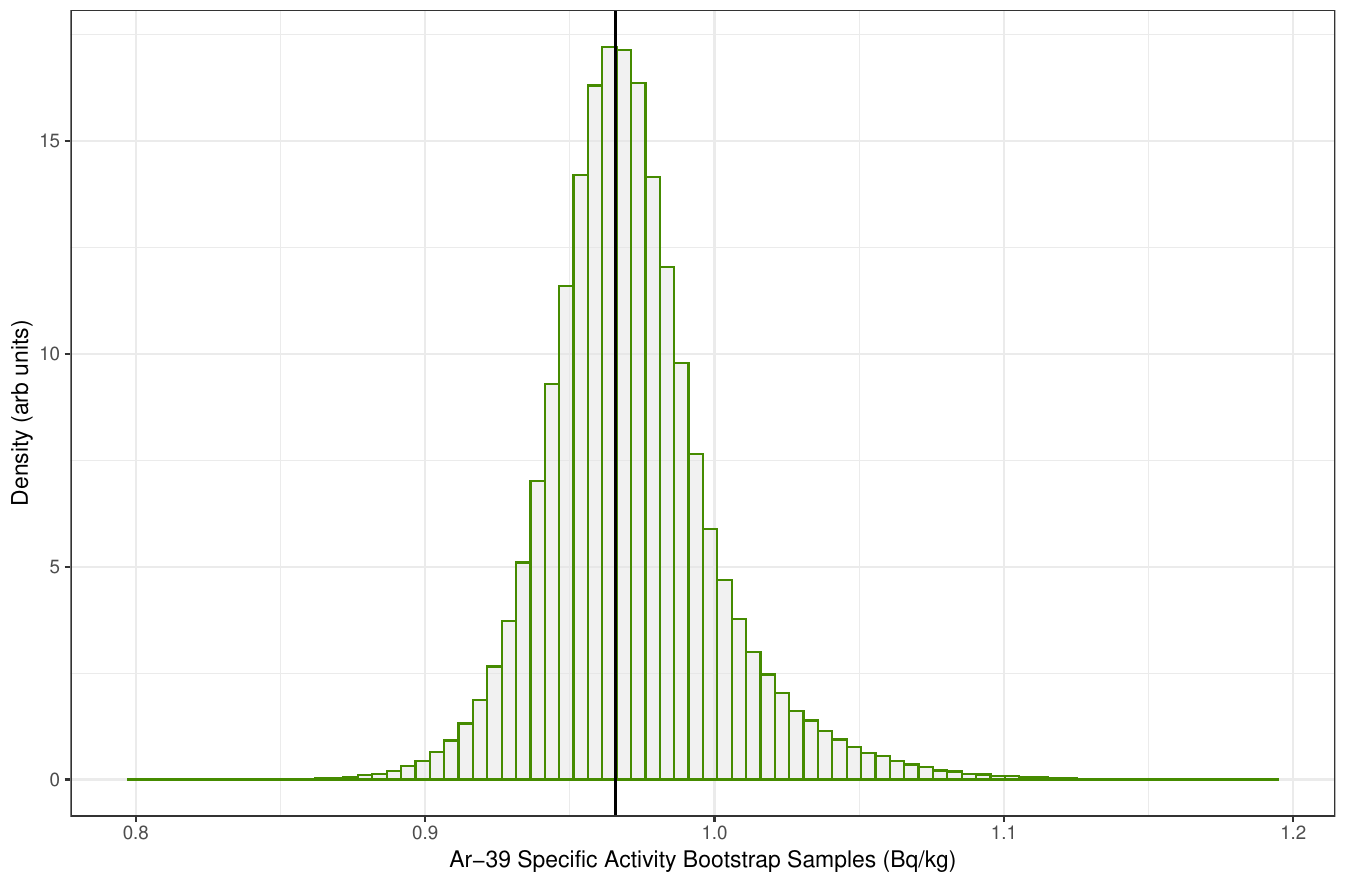}
	\caption{Histogram showing bootstrap samples for the specific activity of $^{39}$Ar obtained from underground measurements. The thin vertical line indicates the MFV (0.966) for the dataset.}
	\label{fig:Ar39SA_HPB_MFV}
\end{figure}

By applying nonparametric bootstrapping with the MFV to estimate the specific activity of $^{39}$Ar from the underground data, the result is 
$ SA_{\text{MFV}}(\text{NP}) = 0.966^{+0.010}_{-0.018} \, \, \text{Bq/kg$_{\text{atmAr} } $} $~\cite{golovkoApplicationMostFrequent2023}. 
This corresponds to a 68.27\% confidence interval of [0.948, 0.976], and a 95.45\% confidence  interval of [0.934, 0.989].
When HPB with MFV is used, the result is $ SA_{\text{MFV}}(\text{HPB}) = 0.966^{+0.027}_{-0.020} \, \, \text{Bq/kg$_{\text{atmAr} } $} $, giving a 68.27\% confidence interval of [0.946, 0.993] and a 95.45\% confidence interval of [0.921, 1.029]. 
To achieve these confidence intervals, 400,000 bootstrap samples were used along with the percentile method~\cite{puth2015variety,mokhtar2023confidence}. This number of bootstrap samples was chosen to ensure that the absolute percent difference in the fitted values for means and their uncertainties was less than 1\% for each element in the dataset when sampling with replacement from the original dataset. Figure~\ref{fig:Ar39SA_HPB_MFV} shows the histogram of all the bootstrap samples created from the original dataset via HPB.

To illustrate the difference between HPB and parametric bootstrapping in estimating the confidence interval for a small dataset, let us assume that the data in Table~\ref{tab:SA_Ar39} follow a normal distribution with a mean value of 0.974 $ \text{Bq/kg$_{\text{atmAr} } $} $. Using parametric bootstrapping with a Gaussian or normal distribution to estimate the confidence interval for the specific activity of $^{39}$Ar from the underground data, we obtain a 68.27\% confidence interval of [0.962, 0.985]~$ \text{Bq/kg$_{\text{atmAr} } $} $ and a 95.45\% confidence interval of [0.951, 0.996]~$ \text{Bq/kg$_{\text{atmAr} } $} $. The result from parametric bootstrapping with Gaussian is $ SA_{\text{Mean}}(\text{PB}) = 0.974^{+0.011}_{-0.011} \, \, \text{Bq/kg$_{\text{atmAr} } $} $. This indicates a 68.27\% confidence interval, which corresponds to a 1$\sigma$ uncertainty.

Using the weighted average method by PDG the specific activity of $^{39}$Ar from underground measurement results is  $ SA_{\text{WA}}(\text{PDG}) = 0.966^{+0.017}_{-0.017} \, \, \text{Bq/kg$_{\text{atmAr} } $} $. The chi-square value is 0.43. Therefore, there is no need to apply a scaling factor to the weighted average uncertainty. 

The central value for the specific activity of $^{39}$Ar is identical whether determined via the MFV method or the weighted average, although the uncertainties differ. The smallest asymmetric confidence interval comes from nonparametric bootstrapping~\cite{golovkoApplicationMostFrequent2023}, highlighting the need for more independent measurements. HPB provides the most cautious estimates of the asymmetric confidence interval because it considers the uncertainty of each data point. In contrast, the weighted average gives a symmetric confidence interval. On the other hand, if we assume that the underlying distribution that describes the dataset in Table~\ref{tab:SA_Ar39} is symmetrical and use Equation~\ref{Eq:sig_M} to estimate the uncertainty for the MFV, we obtain a value of $\pm 0.004  \, \, \text{Bq/kg$_{\text{atmAr} } $}$, which is even smaller than the uncertainty of the weighted average ($\pm 0.017  \, \, \text{Bq/kg$_{\text{atmAr} } $}$).

\section*{Previous $^{97}$R\lowercase{u} Half-Life Measurements}
\label{sec:hisRu97}

Table~\ref{tab:Ru97HL} summarizes all known half-life measurements to date for the $^{97}$Ru isotope to the best of the author’s knowledge. In the table, we have also included historical $^{97}$Ru half-life data that, according to the selection rules of  PDG~\cite{10.1093/ptep/ptaa104}, would not be considered for the unconstrained averaging procedure. Nuclear data sheets for $^{97}$Ru half-life provide an average value of 2.83(23)~days, which is a weighted average of 2.79(3)~days~\cite{KOBAYASHI1998367}, 2.88(4)~days~\cite{katcoff1958branching}, and 2.9(1)~days~\cite{cretzu1966neue}. Data available before November 19, 2009, were taken into account in the nuclear data sheets for estimating the half-life of $^{97}$Ru. It is surprising that the study by Goodwin et al.~\cite{PhysRevC.80.045501} on the $^{97}$Ru half-life was not included in the weighted average, even though it was published on October 16, 2009.

Interestingly, the half-life of $^{97}$Ru was measured during four runs by observing the decay of the photopeak areas of the 215.8 keV and 324.3 keV $\gamma$-rays over approximately 15 days, resulting in a value of 2.79(3) days~\cite{KOBAYASHI1998367}, which was included in the weighted average. On the other hand, an earlier study by the same group measured the decay of $^{97}$Ru over 30 days during three runs by examining the decay of  216~keV $\gamma$-ray photopeak areas and reported a $^{97}$Ru half-life of 2.791(4)~days~\cite{kobayashi1993half}. However, this earlier measurement was not included in the weighted average.

\begin{figure*}[t]
	\centering
	\includegraphics[width=0.9\textwidth]{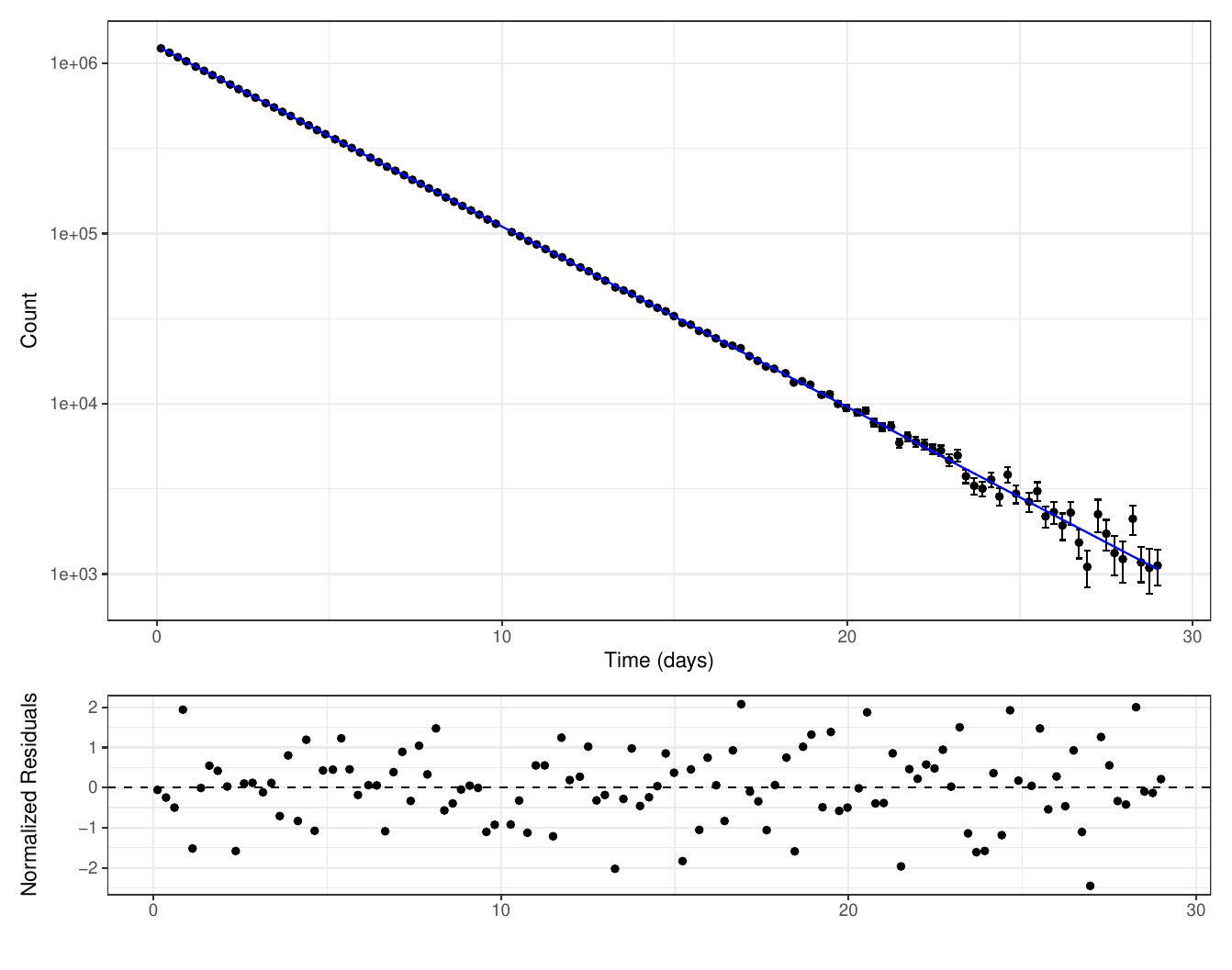}
	\caption{Decay of $^{97}$Ru in ruthenium at 19K. Experimental data are shown as dots along with error bars. The straight line represents a fit to these data using the R fitting algorithm. Normalized residuals are shown at the bottom of the figure. Out of 116 data points, 80 fall within $\pm$1, and 112 fall within $\pm$2. The normalized residuals seem to be spread out randomly.}
	\label{fig:R_Ru97_HL}
\end{figure*}

Szegedi et al.~\cite{szegedi2020high} recently measured the half-life of $^{97}$Ru. They reported statistical uncertainties for only three different alpha irradiation energies used to produce $^{97}$Ru. Moreover, by using the most accurate independent measurements of the $^{97}$Ru half-life, they were able to estimate the systematic uncertainties related to the dead time in the data acquisition system as 0.14\%. Considering this information and incorporating the scaling factor based on the reduced chi-square results for statistical uncertainties, the updated uncertainties for the half-life of $^{97}$Ru are presented in Table~\ref{tab:Ru97HL}. For  uncertainty, we added the statistical and systematic uncertainties in quadrature and used this combined error for the half-life of $^{97}$Ru.

In this work, we aim to examine how different fitting algorithms influence the fit result and its statistical uncertainty. For this purpose, we used the original low-temperature (19K) $^{97}$Ru decay data as presented in~\cite{PhysRevC.80.045501, HARDY20101550, goodwin2013can}. We employed the ROOT and R fitting algorithms.

We configured ROOT~\cite{brun1997root} to use the Simplex~\cite{Nelder1965,cho2016optimass} algorithm from the Minuit2 library (an updated C++ version of Minuit~\cite{JAMES1975343} originally written in FORTRAN) for data fitting. By setting Simplex as the default option, we guaranteed a reliable and efficient fitting process. Unlike the default MIGRAD~\cite{davidon1991variable,cho2016optimass} algorithm in Minuit, which uses derivatives, Simplex operates without  them~\cite{Nelder1965}. This configuration improves the statistical uncertainty estimation of our model fitting, yielding 2.8383(9)~days compared to the previously published result for the same 19K dataset, 2.8382(13)\footnote{The discrepancy in the last digit of the central value is because of a partial access to  the $^{97}$Ru original  half-life data.} days. The statistical uncertainty with the Simplex algorithm is more than 44\% lower than  the one achieved with MIGRAD. 

The $\chi^2$ value for the new fit is 99.56, and the number of data points used is 116. This implies that the reduced $\chi^2$ ($99.56/114$) is less than one, indicating that there is no need to apply a scaling factor to the statistical error.

The values presented in Table~\ref{tab:Ru97HL} for the 19K $^{97}$Ru dataset also include the systematic uncertainty linked to the minor adjustment for residual dead-time losses. Further information on this topic can be found in~\cite{goodwin2013can}.

\begin{table}[H]
	\centering
	\caption{$^{97}$Ru half-life measurements. Uncertainty is expressed as $\pm1\sigma$, representing an approximately a 68.27\% confidence interval. We underlined the values that were used to calculate the weighted average of 2.83(23) days for the half-life of $^{97}$Ru, as adopted in the nuclear data sheets for $A=97$~\cite{NICA2010525}.}
	\label{tab:Ru97HL} 
	\begin{tabular}{cccr}
		\toprule
		\begin{tabular}{c}
			Half-life, \\
			days
		\end{tabular}
		& \begin{tabular}{c}
			Uncertainty, \\
			days
		\end{tabular}
		& Reference & Year \\ 
		\midrule
		2.8000 & 0.3000 & \cite{PhysRev.70.778} & 1946 \\ 
		2.8000 & 0.1000 & \cite{mockPhotoInducedReactions201948} & 1948 \\ 
		2.4400 & NA$^a$ & \cite{PhysRev.100.188} & 1955 \\ 
		\underline{2.8800} & \underline{0.0400} & \cite{katcoff1958branching} & 1958 \\ 
		\underline{2.9000} & \underline{0.1000} & \cite{cretzu1966neue} & 1966 \\ 
		2.8390 & 0.0060 & \cite{Silvester1979} & 1979 \\ 
		2.7910 & 0.0040 & \cite{kobayashi1993half} & 1993 \\ 
		\underline{2.7900} & \underline{0.0300} & \cite{KOBAYASHI1998367} & 1998 \\ 
		2.8370 & 0.0014 & \cite{PhysRevC.80.045501} & 2009 \\ 
		2.8382 & 0.0014$^b$ & \cite{PhysRevC.80.045501} & 2009 \\ 
		2.8160 & 0.0610$^c$ & \cite{larson2013neutron} & 2013 \\ 
		2.8360 & 0.0070$^d$ & \cite{lindstromHalflives90mY97Ru2014} & 2014 \\ 
		2.8387 & 0.0049$^e$ & \cite{szegedi2020high} & 2020 \\ 
		2.8422 & 0.0042$^e$ & \cite{szegedi2020high} & 2020 \\ 
		2.8411 & 0.0040$^e$ & \cite{szegedi2020high} & 2020 \\ 
		\bottomrule
	\end{tabular}
	
	\vspace{0.1cm} 
	
	\begin{minipage}{0.6\columnwidth}
		\raggedright 
		\footnotesize
		$^a$This half-life of the $^{97}$Ru activity was followed with a scintillation spectrometer through more than eight octaves; however, no error was provided.\\
		$^b$This measurement was taken at 19K in ruthenium.\\
		$^c$This half-life measurement of $^{97}$Ru was presented as part of a doctoral thesis. \\
		$^d$This quoted error was recalculated from the value given in the original work, which was based on an approximately 95\% confidence interval. \\
		$^e$This quoted uncertainty includes a systematic error of 0.14\%, which is related to the minor adjustment for dead-time losses.
	\end{minipage}
\end{table}

The decay function model used in the fit is the decay of radioactive $^{97}$Ru over time. It is defined as follows:
\begin{equation}
	N(t) = N_0 e^{-\lambda t}
	\label{eqn:decay}
\end{equation}
where
\( t \) represents time,
\( N_{0} \) denotes the initial amount of $^{97}$Ru, and
\( \lambda \) is the decay constant.
The half-life of $^{97}$Ru, denoted \( T_{1/2} \), is related to the decay constant \( \lambda \) through the following relationship:
\begin{equation}
	\lambda = \frac{\ln(2)}{T_{1/2}}
	\label{eqn:decay_cnst}
\end{equation}

We wanted to investigate how the different minimization software programs estimate the half-life of $^{97}$Ru via the 19K dataset. We want to check whether the application of other minimization algorithms  affects the fit results and their uncertainties. For that, we adopt noncommercial data analysis tools such as R \cite{rcoreteamLanguageEnvironmentStatistical2022}. 
R is a programming language that is open-source and free. It was created in 1993 by statisticians for statisticians. R is well regarded for its ability to perform statistical computations effectively \cite{suttonStatisticsSlamDunk2024}. Furthermore, R is embraced by a wide scientific community and subjected to rigorous verification and validation processes beyond those typically seen in the smaller nuclear physics community. This ensures its reliability and accuracy for data analysis.

The 19K dataset for the decay of $^{97}$Ru in ruthenium was fitted to an exponential function (refer to Equations~\ref{eqn:decay} and \ref{eqn:decay_cnst}) via a nonlinear least-squares algorithm~\cite{elzhovMinpackLmInterface2023} to minimize \(\chi^2\). This was implemented with the Levenberg--Marquardt method~\cite{more1978levenberg} in R to determine the half-life $T_{1/2}$ and the initial amount of $^{97}$Ru. The weight of each data point was also considered during the minimization process. The results of our fitting process are shown in Figure~\ref{fig:R_Ru97_HL}. 

The decay data fit a single exponential model well, yielding a half-life for $^{97}$Ru of 2.8383(12)~days, where the uncertainty represents a statistical error at the 68.3\% confidence level. This result is very similar to that obtained with the Minuit2/Simplex algorithm, which is 2.8383(9)~days, and a new ROOT fit demonstrates a  smaller statistical uncertainty. We computed  \(\chi^2=99.56\) and compared it with $N-2=114$, which is the expected \(\chi^2\) value if the measurements follow a Gaussian distribution. Therefore, the reduced \(\chi^2\) value is 0.8733, which is less than one. Therefore, there is no requirement for a scale factor to be applied to the statistical uncertainty.

Following the recommendation of the PDG, we applied unconstrained averaging for uncorrelated measurements, assuming that they follow a Gaussian (normal) distribution. Despite the small dataset presented in Table~\ref{tab:Ru97HL}, it is still possible to test these data for normality. When dealing with small sample sizes, the Shapiro--Wilk test~\cite{shapiro1965analysis} is often regarded as the most powerful and reliable test for normality. This test is powerful even with very small sample sizes (for example, $n=3$) to identify deviations from normality~\cite{kim2019more}. For a sample size of $n=10$, the Shapiro--Wilk test outperforms the Kolmogorov--Smirnov, Anderson--Darling, and Lilliefors tests in performance~\cite{razali2011power}. The Anderson--Darling test~\cite{anderson1952asymptotic}, however, shows performance close to that of the Shapiro--Wilk test. Nonetheless, even if the Shapiro--Wilk test confirms normality with a small dataset, it does not necessarily mean that the data originated from a normal distribution; the test might simply lack the power to detect nonnormality. 

\begin{figure}[t]
	\centering
	\includegraphics[width=0.99\columnwidth]{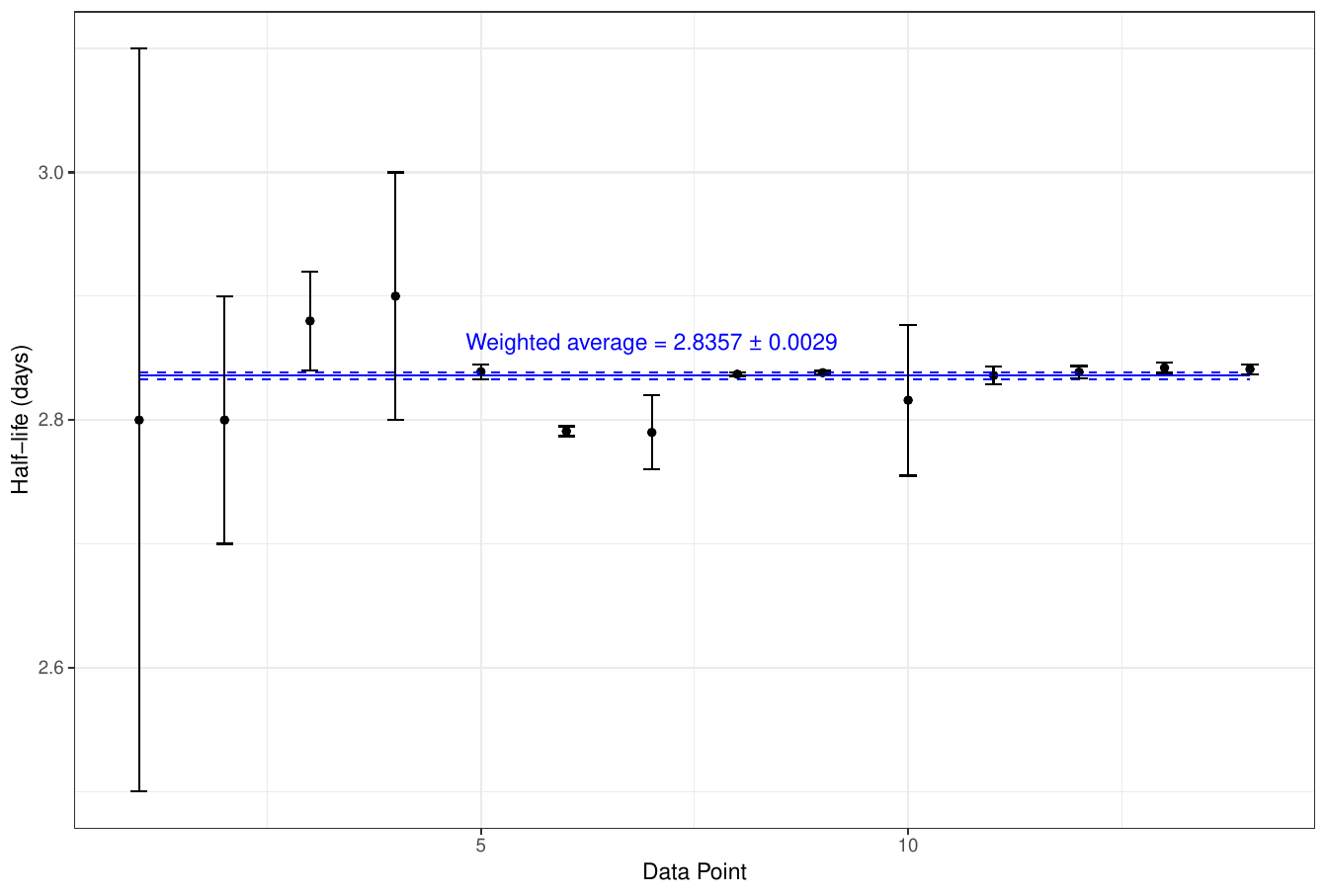}
	\caption{Half-life measurements for $^{97}$Ru, along with their uncertainties (see Table~\ref{tab:Ru97HL}). Only the data with known uncertainties are included. The weighted average (2.8357) and its corresponding uncertainty are also shown. The uncertainty of the weighted average was increased by a scaling factor of 3.26, as explained in the `\nameref{subs:wam}' subsection. }
	\label{fig:Ru97_HL_W.m}
\end{figure}

The Shapiro--Wilk test for the dataset shown in Table~\ref{tab:Ru97HL} (excluding the 1955 data without uncertainty) yields a p-value ($p=0.10$) greater than 0.05. A p-value greater than 0.05 indicates that we fail to reject the null hypothesis that the data follow a normal distribution. Similarly, the Anderson--Darling test's p-value ($p=0.06$) is just above 0.05. Although this suggests some evidence against normality, it is not strong enough to definitively reject the normality assumption. If we include the 1955 data, the results from the Shapiro--Wilk and Anderson--Darling tests all suggest that the half-life data do not follow a normal distribution, which is supported by significant p-values (close to 0). p-values close to 0 indicate strong evidence against the null hypothesis, suggesting that the data are not normally distributed. Therefore, the 1955 $^{97}$Ru half-life data might be considered outliers in the dataset. 

Figure~\ref{fig:Ru97_HL_W.m} shows the measured half-life of $^{97}$Ru, along with its known uncertainty (refer to Table~\ref{tab:Ru97HL}). The 1955 half-life value would be outside the range displayed in Figure~\ref{fig:Ru97_HL_W.m}. The chi-square value for the data in Figure~\ref{fig:Ru97_HL_W.m} is 138.033, so the uncertainty in the weighted average of the $^{97}$Ru half-life of $2.8357\pm0.0029$~days was adjusted with a scaling factor as explained in the subsection `\nameref{subs:wam}.' The MFV for the half-life of $^{97}$Ru is $2.8385 \pm 0.0005$~days, with the uncertainty calculated via Equation~\ref{Eq:sig_M}. 

As an exercise, if we include the half-life value from 1955 in the dataset for the MFV estimation, the value and its uncertainty remain the same at $2.8385 \pm 0.0005$~days. This demonstrates an advantage of  MFV estimation, as it is not affected by outliers in the dataset. Alternatively, one can assume the uncertainty in the 1955 measurement to be the last digit of the central value, or $2.44\pm0.01$~days. This assumption permits the use of the weighted average estimation for all the half-life measurements of $^{97}$Ru presented in Table~\ref{tab:Ru97HL}. This approach results in a value of $2.8327\pm0.0096$~days, with the uncertainty adjusted by a scaling factor of 10.99. Both the central value and its uncertainty clearly changed in the weighted average estimation.

\begin{figure}[t]
	\centering
	\includegraphics[width=0.99\columnwidth]{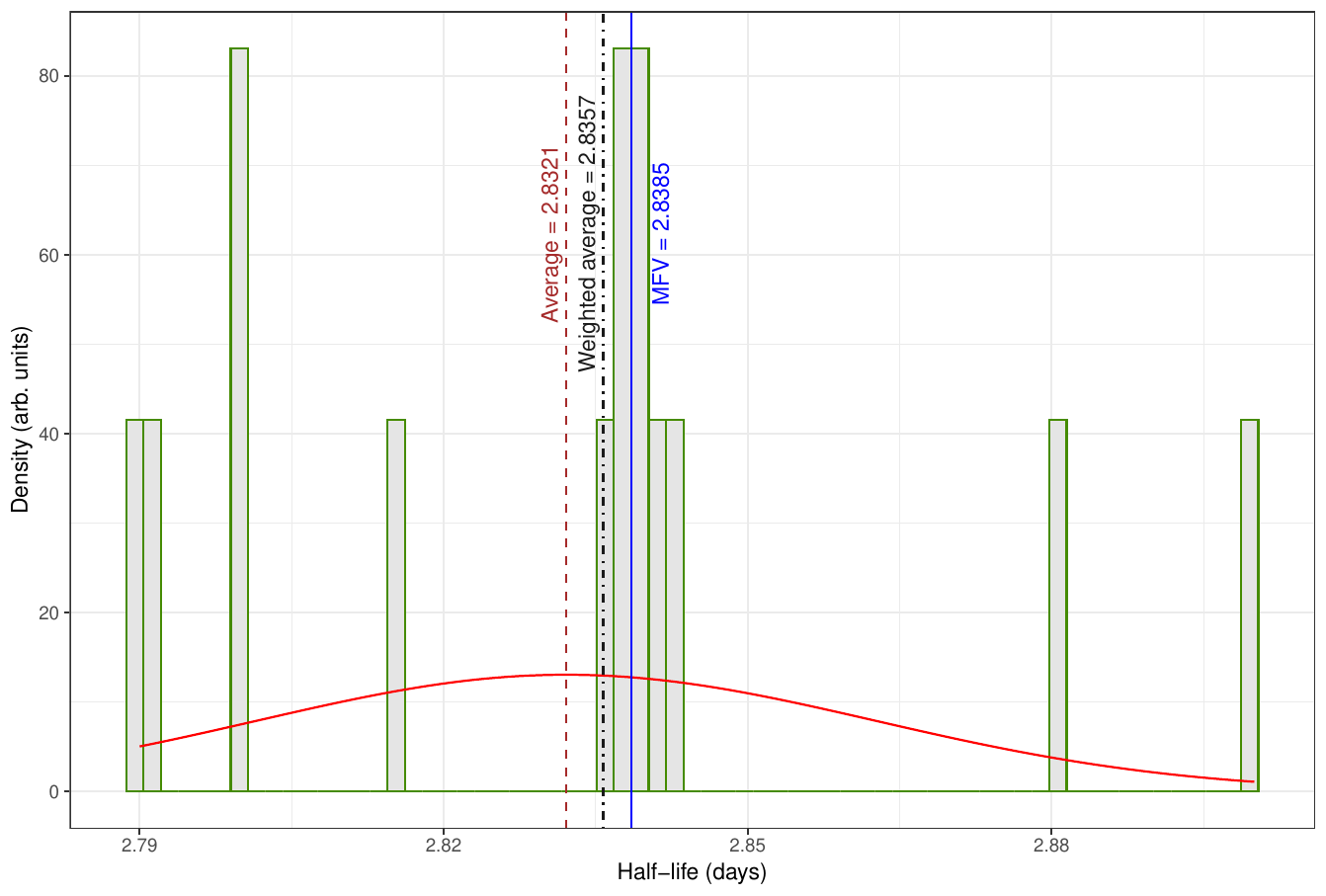}
	\caption{Histogram of $^{97}$Ru half-life measurements (refer to Table~\ref{tab:Ru97HL}), which includes only the data with known uncertainties. The weighted average value (2.8357) is represented by a vertical dot-dashed line,  the MFV (2.8385) is indicated by a vertical solid line, and the average (2.8321) is shown by a vertical dashed line. The normal data distribution, which describes the original dataset and is used for parametric bootstrapping (see section `\nameref{sec:disc}'), is also presented.} 
	\label{fig:Ru97his}
\end{figure}

\begin{figure}[t]
	\centering
	\includegraphics[width=0.99\columnwidth]{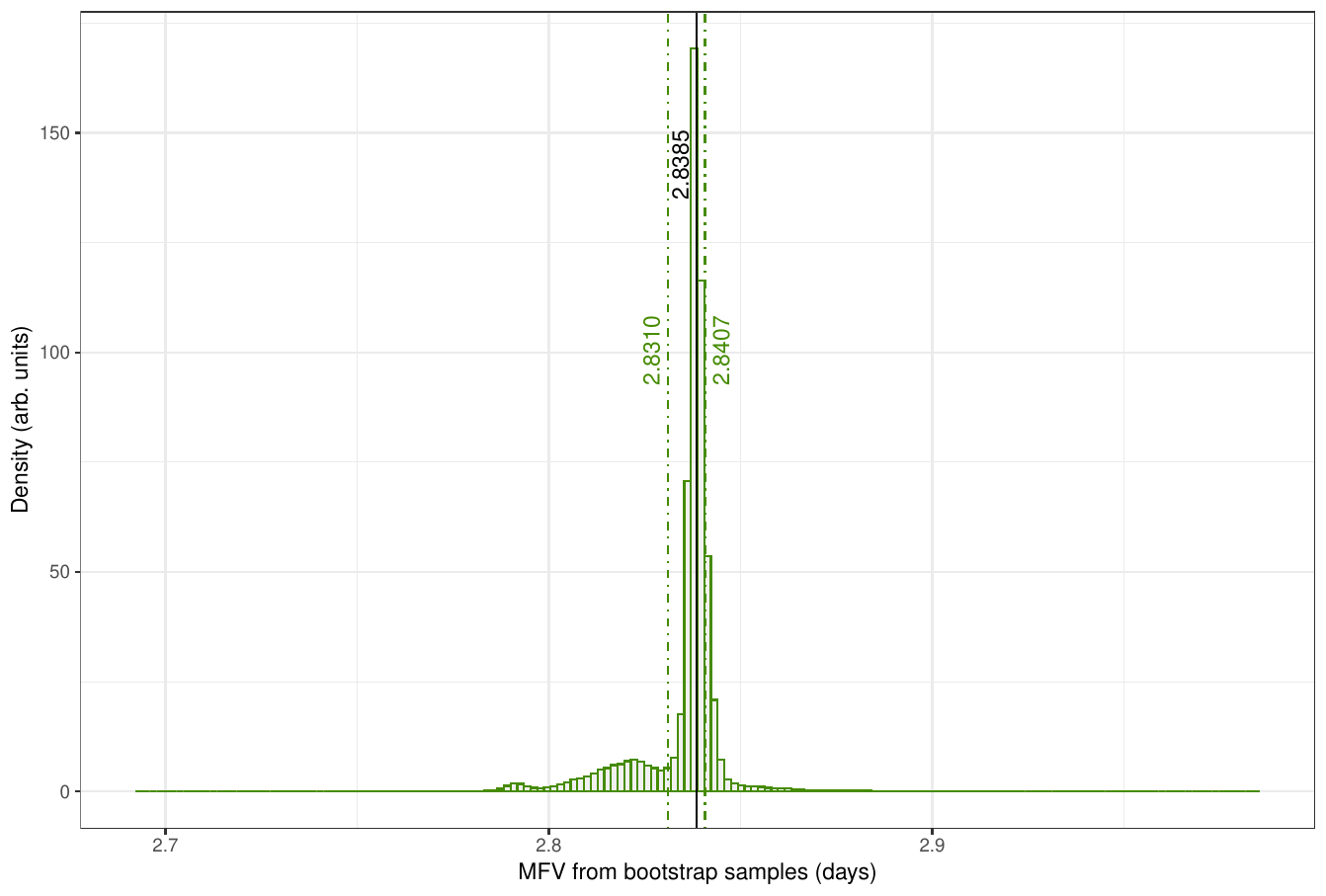}
	\caption{Histogram displaying all the MFVs from the bootstrap samples (see Figure~\ref{fig:Ru97his}). The vertical solid line indicates the MFV (2.8385), and the dash-and-dot lines present the 68.27\% confidence interval calculated with HPB via the percentile method.}
	\label{fig:Ru97his_boot}
\end{figure}

Figure~\ref{fig:Ru97his} presents the histogram of $^{97}$Ru half-life measurements, including only historical data with known uncertainties (refer to Table~\ref{tab:Ru97HL}). As discussed in the subsection `\nameref{subs:boot},' for the small dataset depicted in Figure~\ref{fig:Ru97his}, it is more appropriate to use the HPB method to estimate the confidence interval of the $^{97}$Ru half-life. This method accounts for the uncertainty of each individual measurement and does not make assumptions about the underlying distribution of the data. When HPB with MFV was used, the $^{97}$Ru half-life result was $ T_{1/2\text{,MFV}}(\text{HPB}) = 2.8385^{+0.0022}_{-0.0075} \, \, \text{days} $. This provides a 68.27\% confidence interval of [2.8310, 2.8407] and a 95.45\% confidence interval of [2.8036, 2.8485]. To generate these intervals, 210,000 bootstrap samples were evaluated via the percentile method~\cite{puth2015variety,mokhtar2023confidence}. This number of bootstrap samples was selected to ensure that the absolute percent difference in the estimated means and their uncertainties was less than 1\% for each element in the bootstrap dataset when sampling with replacement from the original dataset. Figure~\ref{fig:Ru97his_boot} presents the histogram of all the MFVs from bootstrap samples generated from the original dataset shown in Figure~\ref{fig:Ru97his} via HPB.

\section*{Discussion}\label{sec:disc}

In the previous section, we described the use of the HPB method along with the MFV approach to estimate the half-life of $^{97}$Ru and its uncertainty via a small historical dataset. The result is as follows:
\begin{equation}
	T_{1/2\text{,MFV}}(\text{HPB}) = 2.8385^{+0.0022}_{-0.0075} \, \, \text{days},
	\label{eqn:T_1_2}
\end{equation}
where the uncertainty represents a 68.27\% confidence interval.
Let us look into how the confidence interval changes when we use the nonparametric bootstrap technique combined with the MFV approach, and apply the percentile method. The resulting value is $ T_{1/2\text{,MFV}}(\text{NP}) = 2.8385^{+0.0005}_{-0.0162} \, \, \text{days} $. This uncertainty represents a 68.27\% confidence interval of [2.8223, 2.8390], whereas the 95.45\% confidence interval is [2.8085, 2.8411]. Even though the central values are identical, the confidence intervals differ because the nonparametric method does not consider the uncertainty of each individual data point in the dataset. 

On the other hand, the nonparametric bootstrap method is relatively quicker because it does not require reaching an adequately large number of bootstrap samples to ensure that the absolute percent difference in the estimated means and their uncertainties is less than 1\% (or better) for each element. This requirement is crucial to avoid adding extra sources of uncertainty when assessing the spread of the original dataset. Thus, the HPB method is more computationally demanding than the nonparametric bootstrapping is. However, for large datasets,  nonparametric bootstrapping effectively estimates confidence intervals~\cite{efron1994introduction} without requiring knowledge of the original dataset's distribution function. One more benefit of the HPB method is that, in theory, it can handle nonsymmetrical uncertainties (see, for example, Equation~\ref{eqn:T_1_2}) for each individual element in the dataset, provided that an appropriate distribution function is known.

As we previously mentioned, the small $^{97}$Ru half-life dataset presented in Table~\ref{tab:Ru97HL} can also be analyzed via a parametric bootstrap, assuming that the underlying distribution is known. In most cases, it is presumed to be normal or Gaussian. We assume that the data with known uncertainties in Table~\ref{tab:Ru97HL} follow a normal distribution with a mean value of 2.8321~days. Using parametric bootstrapping with a Gaussian function to estimate the confidence interval for the $^{97}$Ru half-life from historical data, we find a 68.27\% confidence interval of [2.8239, 2.8402]~days and a 95.45\% confidence interval of [2.8158, 2.8484]~days. The result from parametric bootstrapping with a Gaussian distribution (see Figure~\ref{fig:Ru97his}) is $ T_{1/2\text{,Mean}}(\text{PB}) = 2.8321^{+0.0081}_{-0.0082} \, \text{days} $. This shows that the 68.27\% confidence interval corresponds to a 1$\sigma$ uncertainty. 

Notably, the results of bootstrapping with a Gaussian distribution lack robustness and can be significantly altered by just one outlier, specifically the data from 1955. Indeed, including the 1955 data in the set reveals that the underlying distribution does not pass the normality test. As a result, this distribution for all the data from Table~\ref{tab:Ru97HL} should not be modeled via the Gaussian (normal) distribution. On the other hand, the MFV is based on the minimization of information loss~\cite{steinerMostFrequentValue1988} and is very resistant to the presence of outliers in the dataset. Additionally, the HPB method does not assume any specific distribution for the data. It is beneficial to use the MFV along with the HPB to analyze small datasets. 

It is also known that using the PDG weighted average procedure with a scaling factor  often favors high-precision measurements, at least as much as the least-squares unconstrained average does~\cite{10.1093/ptep/ptaa104}. Therefore, it places significant emphasis on selecting the data carefully. When combining historical datasets, one should decide which data to include in the set. The MFV algorithm tries to incorporate inconsistent data into a meaningful central value, focusing on where most of the data are concentrated. 

There are two basic situations for historical datasets with outliers: (a) data points that are far from most of the data have errors (such as unreported mistakes), and (b) the primary set of data is inaccurate. PDG provides a historical perspective of the values of a few particle properties tabulated in~\cite{10.1093/ptep/ptaa104} as a function of the date of publication. It is fascinating to observe the evolution of measurements for neutron lifetime. Recent findings demonstrate varying degrees of accuracy across different measurement techniques such as beam, bottle, or space~\cite{zhang2022mfv} used to measure neutron lifetime. The use of the MFV algorithm in combination with nonparametric bootstrapping to analyze historical neutron lifetime data is especially interesting. However, any analysis algorithm will not work well if the main set of data is incorrect. The MFV method  also fails in this situation because most of the data are from the primary set. 

\section*{Conclusions} \label{sec:conclusion}

A hybrid parametric bootstrap method was created to analyze small datasets. This technique can be used on small datasets without prior knowledge of the data's probability distribution. Additionally, HPB considers the uncertainty in each individual element of the dataset. When used alongside the most frequent value method, HPB allows for a highly confident estimation of the 1$\sigma$ and 2$\sigma$ confidence intervals. HPB has been compared with the parametric bootstrap approach, which can also be used on small datasets but requires prior knowledge of the underlying probability distribution of the data in the original dataset. 

The HPB and MFV methods were used to estimate the half-life of $^{97}$Ru, resulting in a value of $ T_{1/2\text{,MFV}}(\text{HPB}) = 2.8385^{+0.0022}_{-0.0075} \, \text{days} $. This corresponds to a 68.27\% (1$\sigma$) confidence interval of [2.8310, 2.8407] days and a 95.45\% (2$\sigma$) confidence interval of [2.8036, 2.8485] days. This value is at least 30 times more precise than the currently accepted half-life of $^{97}$Ru listed in the nuclear data sheets, which is $ T_{1/2} = 2.83^{+0.23}_{-0.23} \, \text{days} $~\cite{NICA2010525}. Furthermore, it is comparable in precision to the most accurate measurement available~\cite{PhysRevC.80.045501}. The HPB method was also compared with the nonparametric and parametric bootstrap methods for estimating the confidence intervals of the $^{97}$Ru half-life. 

Using the HPB method along with the MFV approach, we reassessed a small set of four element data combined from the specific activity of $^{39}$Ar in underground measurements. The result obtained is $ SA_{\text{MFV}}(\text{HPB}) = 0.966^{+0.027}_{-0.020} \, \, \text{Bq/kg$_{\text{atmAr} } $} $. This result offers a 68.27\% confidence interval of [0.946, 0.993] and a 95.45\% confidence interval of [0.921, 1.029]. Once more, the HPB method was compared with the nonparametric and parametric bootstrap methods for estimating the confidence intervals of the specific activity of $^{39}$Ar in underground measurements.


\section*{Data and code availability}

\begin{itemize}
    \item This manuscript has associated data available in the following repository: \\ \href{https://osf.io/a5cvx/}{https://osf.io/a5cvx/}~\cite{golovkoDataEstimation97Ru2024}.
    \item Our source code to analyze ``cold'' $^{97}$Ru half-life data is available at the following repository:  \href{https://osf.io/a5cvx/}{https://osf.io/a5cvx/}~\cite{golovkoDataEstimation97Ru2024}.
    \item This paper has an associated Wiki page at the following repository: \\ \href{https://osf.io/a5cvx/}{https://osf.io/a5cvx/}~\cite{golovkoDataEstimation97Ru2024}.    
    
\end{itemize}


\section*{Acknowledgments}

I would like to express my sincere gratitude to Prof. Dr. John Goodwin for sharing ``cold'' $^{97}$Ru half-life data. I am truly grateful for the support and assistance provided by Maria Filimonova. I would also like to extend my appreciation to the management and staff at Canadian Nuclear Laboratories for fostering an enabling environment for this study, with special mention of Genevieve Hamilton and David Yuke. I am indebted to Helena Rummens for her meticulous editing of the paper. 

\section*{Author contributions}


The author  contributed solely to the conception, design, research, writing, and revision of this manuscript. All aspects of the study were independently carried out by the author.




\section*{Declaration of generative AI and AI-assisted technologies}

During the preparation of this work the author used ChatGPT  to check the language of the manuscript. After using this tool, the author reviewed and edited the content as needed and takes full responsibility for the content of the publication.

\bigskip

\end{document}